\documentclass[12pt]{article}
\usepackage{amsmath}
\usepackage{graphicx,psfrag,epsf}
\usepackage{enumerate}
\usepackage{natbib}
\usepackage{url} 

\usepackage{booktabs,threeparttable}
\usepackage{bbm}
\usepackage{mathtools}
\usepackage{amsthm}

\usepackage{bm}
\usepackage[ruled,vlined]{algorithm2e}
\providecommand{\SetAlgoLined}{\SetLine}
\newcommand{\indep}{\perp \!\!\! \perp}
\usepackage{graphicx}
\graphicspath{ {./images/} }
\usepackage{xcolor}
  \usepackage{etoolbox}
\newcommand\red[1]{{\color{black}#1}}

\newcommand\revisetwo[1]{{\color{black}#1}}
\newcommand\revisethree[1]{{\color{black}#1}}

  \newcommand{\Lower}[1]{\smash{\lower 1.5ex \hbox{#1}}}

\def\bfg{{\ensuremath{\bf g}}}
\def\bfh{{\ensuremath{\bf h}}}

\def\bfH{{\ensuremath{\bf H}}}

\def\bfZ{{\ensuremath{\bf Z}}}
\def\bfzero{{\ensuremath{\bf 0}}}

\def\bftheta{{\ensuremath\boldsymbol{\theta}}}
\def\bfxi{{\ensuremath\boldsymbol{\xi}}}

\def\bfalpha{{\ensuremath\boldsymbol{\alpha}}}

\begin{document}

\def\spacingset#1{\renewcommand{\baselinestretch}%
{#1}\small\normalsize} \spacingset{1}

\author{}
{
  \title{\bf  The Effect of \red{Alcohol intake} on \revisethree{Brain White Matter Microstructural Integrity}: A New {Causal Inference} Framework for Incomplete Phenomic Data}
  \author{Chixiang Chen$^{1}$\thanks{
    \textit{Contact: chixiang.chen@som.umaryland.edu}}, Shuo Chen$^{1}$, Zhenyao Ye$^{1}$, \\
    Xu Shi$^{2}$, Tianzhou Ma$^{3}$, Michelle Shardell$^{1,4}$\hspace{.2cm}\\
    $^{1}$ Division of Biostatistics and Bioinformatics, \\ University of Maryland School of Medicine, Baltimore, U.S.A.\\
    $^{2}$Department of Biostatistics, University of Michigan, Ann Arbor, U.S.A.\\
    $^{3}$Department of Epidemiology and Biostatistics, \\
    University of Maryland, College Park, U.S.A. \\
    $^{4}$ Institute for Genome Sciences, \\ University of Maryland School of Medicine, Baltimore, U.S.A.\\}
  \maketitle
} 


\begin{abstract}

 Although substance use, such as alcohol intake, is known to be associated with cognitive decline during aging, its direct influence on the central nervous system remains incompletely understood. In this study, we investigate the influence of alcohol intake frequency on \revisethree{reduction of brain white matter microstructural integrity in the fornix, a brain region considered a promising marker of age-related microstructural degeneration}, using a large UK Biobank (UKB) cohort with extensive phenomic data reflecting a comprehensive lifestyle profile. Two major challenges arise: 1) potentially nonlinear confounding effects from phenomic variables and 2) a limited proportion of participants with complete phenomic data. To address these challenges, we develop a novel ensemble learning framework {tailored for robust causal inference} and introduce a data integration step to incorporate information from UKB participants with incomplete phenomic data, improving estimation efficiency. Our analysis reveals that daily alcohol intake may \revisethree{significantly reduce fractional anisotropy, a neuroimaging-derived measure of white matter structural integrity, in the fornix} and increase systolic and diastolic blood pressure levels. Moreover, extensive numerical studies demonstrate the superiority of our method over competing approaches in terms of estimation bias, \revisethree{while outcome regression-based estimators may be preferred when minimizing mean squared error is prioritized.}
 
\end{abstract}

\noindent%
\vfill

\spacingset{1.9} 
\section{Introduction}\label{Introduction}

The availability of high-dimensional phenotypic data, referred to as phenomic data, offers new opportunities to study physical and biochemical traits of living organisms \citep{houle2010phenomics}. 
{Our study is motivated by the potential influence of substance use, \red{such as alcohol intake}, on white matter brain aging, as characterized by neuroimaging techniques. Although \red{recent literature has shown that high levels of alcohol intake relate to increased risk of accelerated cognitive decline and all-cause mortality during aging \citep{mende2019alcohol,zhao2023association}, the direct influence of alcohol intake on the central nervous system remains incompletely understood}. \revisethree{Meanwhile, numerous studies have consistently reported strong and replicable associations between aging and brain white matter microstructural integrity, as measured by fractional anisotropy (FA) derived from diffusion-weighted imaging, a more objective indicator of brain aging than cognitive performance measures. Notably, white matter microstructural integrity may explain up to $94\%$ of variance in age-related cognitive decline \citep{lee2024new}. Among the various white matter tracts, the fornix, a central white matter bundle in the limbic system \citep{perez2023alcohol}, is a key anatomical structure involved in memory-related processes. The fornix exhibits pronounced sensitivity to aging-related microstructural degradation \citep{chen2015age} (Figure~\ref{fig:workflow}A), where reductions in fornix FA have been associated with poorer memory performance and are implicated in several neurodegenerative disorders such as Alzheimer's disease \citep{kantarci2014fractional, bangen2021decreased}. Additionally, emerging evidence from both human and animal models suggests a potential negative association between fornix FA and alcohol exposure, as observed through parametric association analyses \citep{zahr2017alcohol,daviet2022associations}. {Motivated by these findings, the primary objective of this study} is to assess the impact of alcohol intake on fornix FA using advanced causal inference approaches.  } 


\begin{figure}
    \centering
    \includegraphics[width=5in]{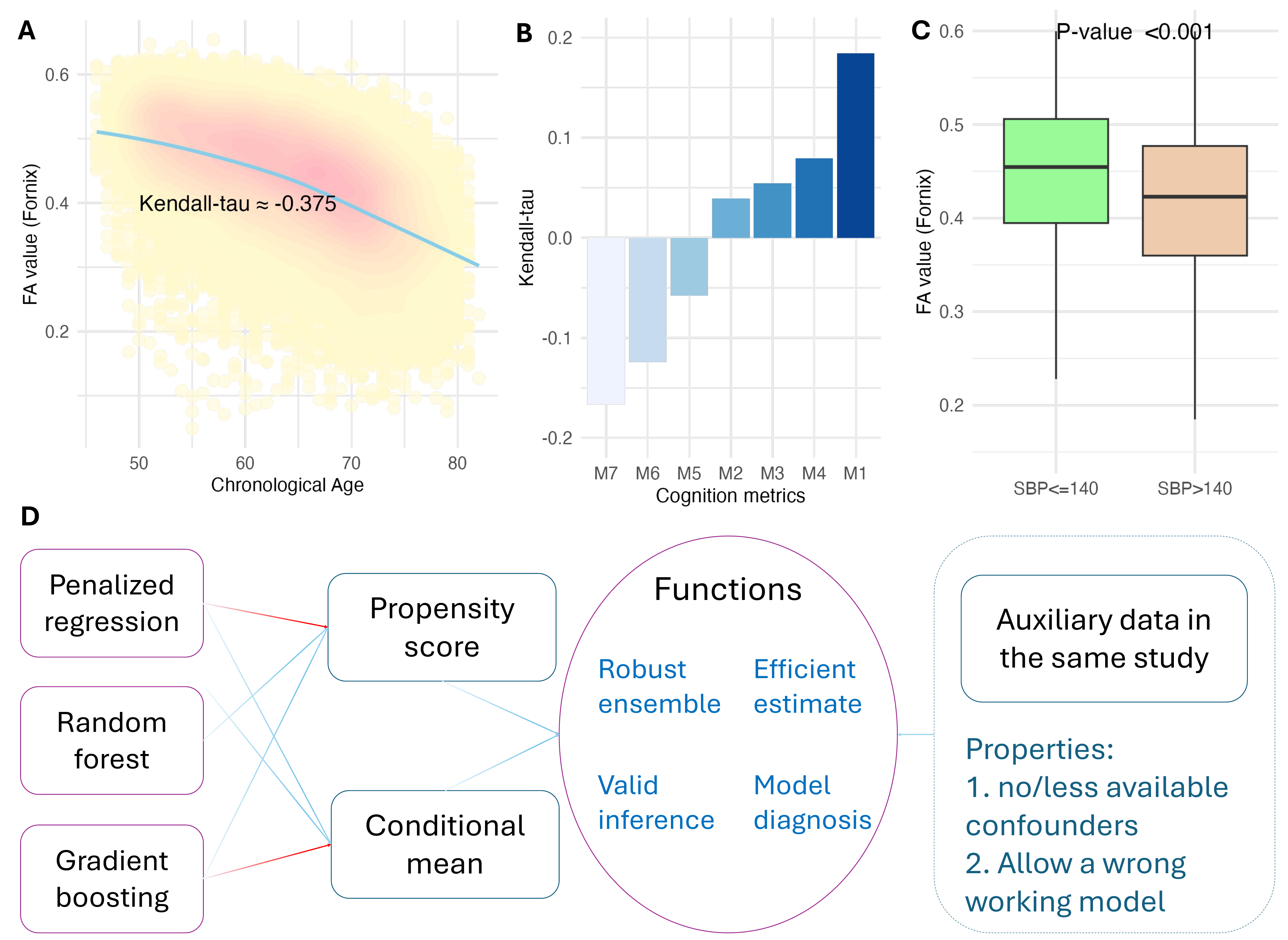}
    \caption{A-C: Fornix FA and aging-associated outcomes. SBP: systolic blood pressure, M1-M7 are cognitive performance measures described in Section 4; D: Methodology overview. }
    \label{fig:workflow}
\end{figure}


{To account for multiple factors that \revisetwo{potentially} confound the relationship between \red{alcohol intake} and fornix FA, we use data from the UK Biobank (UKB), which provides phenomic data with unprecedented breadth and depth \citep{mende2019alcohol}, including measurements that are not jointly present in most previous studies such as multimodal brain imaging, nutrition intake, physical activities, and comprehensive blood measures inclusing metabolomics. 
{These phenomic variables jointly reflect a comprehensive life-style profile.} After applying inclusion/exclusion criteria and operationally defining candidate variables, \revisethree{$3,435$} participants had complete phenomic data for the study of alcohol intake (referred to as the main data), while \revisethree{$21,874$} participants had extensive missing phenomic variables (referred to as the auxiliary data). 
\revisetwo{Missing data arise from the UKB's phased study design, which prioritized breadth of recruitment over uniform data collection. While most participants completed baseline assessments (e.g., questionnaires, physical measurements), only subsets underwent more resource-intensive evaluations and measurements. For example, nuclear magnetic resonance spectroscopy and other assays were performed to characterize plasma biomarkers on stratified samples or subgroups due to logistical, budgetary, and operational constraints. Additional missingness reflects variable consent for specific data uses (e.g., genetic or biomarker analyses), exclusion of measurements due to quality control protocols (e.g., insufficient sample volume or technical errors), and staggered data releases over time \citep{UKB_AboutData}.} Simply merging these two (main and auxiliary) datasets could introduce bias due to missing confounders, whereas relying solely on the main data may result in low statistical efficiency. Therefore, to achieve both unbiased and efficient estimation, we face two key statistical challenges: (1) how to \revisetwo{handle phenomic variables} in the main data, given their unknown and complex confounding effects? (2) \red{how to improve efficiency of estimating the mean potential outcome in the main data, potentially by leveraging information from the larger auxiliary dataset?}}

Substantial contributions have been made to statistical methods that can adjust for observed confounders, including outcome conditional mean imputation \citep{robins1999testing}, (augmented) inverse probability weighting \citep{horvitz1952generalization,robins1994estimation,bang2005doubly}, and matching \citep{rubin1973matching,rubin2006matched,antonelli2018doubly}. Most methods require specification and estimation of nuisance functions, such as the outcome conditional mean and/or propensity score.
 When handling \revisetwo{many} covariates with potentially non-linear confounding effects in estimating nuisance functions, \red{causal machine-learning methods using random forest or gradient boosting are preferred to classic regression and lead to tractable statistical properties} \citep{chernozhukov2018double,yang2023elastic,cui2023estimating}. Despite substantial efforts, a considerable challenge of existing machine-learning-based methods is the need for prior knowledge to choose the best learner, especially given the extensive options among machine learning algorithms. Machine-learning algorithms have a black-box characteristic, and their performance can substantially vary across databases and underlying set-ups. Although advanced computational schemes such as Super Learner \citep{polley2010super} are numerically capable of integrating multiple machine learning algorithms, these methods \revisetwo{are primarily designed for optimizing outcome prediction but may not be well-suited for causal inference, where the goal is to address confounding and balance covariates among exposure groups.} Thus, how to effectively integrate multiple machine-learning models and robustly estimate the population mean potential outcome \revisetwo{remains an open research challenge}.   

On the other hand, 
a promising attempt to improve estimation efficiency is to integrate information from external data or auxiliary records within the data. In recent decades, researchers have employed many different information integration schemes, such as meta-analysis, generalized meta analysis, empirical likelihood methods, constrained maximum likelihood methods, and Bayesian methods with informative priors
\citep{chatterjee2016constrained,yang2019combining,zhang2020generalized,jiang2021elastic,zhai2022data,chen2023efficient,chen2024integrating}. 
 Most of these methods rely on summary information extracted from external sources and have not been applied in the context of causal inference. In our application, we investigate the potential of integrating information from auxiliary data with incomplete phenomic records to enhance the efficiency of estimating the mean potential outcome in the main data, an important and understudied topic in the literature.

To simultaneously and robustly address the challenges of complex confounding and an incomplete confounder set in the auxiliary data, we propose a novel statistical framework incorporating an ensemble learner that ensures both robustness and efficiency (Figure~\ref{fig:workflow}D).
 This learner has several unique advantages over aforementioned methods. First, our learner allows multiple machine learning algorithms to \revisetwo{calibrate imbalanced covariate distributions resulting from non-random exposure assignment}. The resulting estimator is robust in the sense that it is consistent and leads to a desirable convergence rate and tractable statistical properties if at least one algorithm captures the true propensity score and at least one algorithm captures the true conditional mean of the outcome. This ensemble learner does not require prior knowledge about the best algorithm. Second, the proposed learner enhances estimation efficiency by incorporating information from auxiliary data. In our application, the auxiliary data consist of study participants whose phenomic data are only partially observed, largely by design. This setting can be seen as an extreme missing data problem due to the presence of numerous unobserved covariates, where traditional missing-data techniques, such as imputation, may not be suitable. Our proposed integration scheme efficiently and \revisetwo{robustly} uses such auxiliary data to boost the estimation efficiency of the learner, by proposing novel informative scores via empirical likelihood \citep{qin1994empirical}. This novel learner is computationally efficient and flexibly accommodates exposures \revisetwo{with more than two levels}. Its application has yielded several critical insights that enhance our understanding of the effects of alcohol intake frequency on fornix FA. \revisethree{Furthermore, extensive simulation studies demonstrate that our proposed estimators consistently yield low estimation bias, while outcome regression-based estimators are preferred when prioritizing mean squared error (MSE) minimization.} 

The remaining sections of this paper are organized as follows. Section \ref{Notation} introduces the notation. Section \ref{Method} presents the proposed framework for robust causal machine learning and data integration. Section \ref{A real data application} provides a case study on the impact of alcohol intake frequency on the mean potential outcome of fornix FA in the UKB. Section \ref{Simulation} includes numerical evaluations of the proposed estimators. Section \ref{Discussion} discusses and summarizes the methodology and its application. \revisetwo{All theoretical results with technical details}, numerical procedures, and additional real data results and simulations can be found in the Supplementary Material.

\section{Notation}\label{Notation}
For $i=1,\ldots,n$, let $(Y_i,X_i,\bfZ_i^{\scriptscriptstyle \top})^{\scriptscriptstyle \top}$ be the independent and identically distributed main data for subject $i$, where $Y$ is a univariate outcome of interest \revisetwo{(i.e., \revisethree{measured fornix FA})}, \red{$X=x\in\{0,\ldots, L\}$ is a univariate exposure variable with in total $L+1\geq 2$ levels \revisetwo{(e.g., alcohol intake frequencies)}, and $\bfZ$ is a $p$-dimensional vector consisting of observed confounders of the outcome $Y$ and exposure $X$. In addition to the main data, we consider an auxiliary dataset: for $i=n+1,\ldots,N$, let $(Y_i,X_i)^{\scriptscriptstyle \top}$ be the independent and identically distributed auxiliary data. We focus on the scenario in which the raw auxiliary data $Y$ and $X$ are available, while the variables in $\bfZ$ are either completely unobserved or only partially observed. \revisetwo{As discussed in Section \ref{Introduction} and Section \ref{A real data application}, the missing covariates in our case study were due to design and resource constraints, and similar participant profiles were observed between the main and auxiliary cohorts, as shown in Table \ref{tab:demo}}. Therefore, we assume that both datasets follow the same distribution.
Potential violations of this homogeneity assumption will be discussed in Sections \ref{Information integration} and \ref{Discussion}.
Moreover, under the Stable Unit Treatment Value Assumption, let $Y_i(x)$ be the potential outcome of subject $i$ if the exposure status had, possibly contrary to fact, been set to $X=x$. Then the population mean potential outcome is defined as
\begin{equation}\label{ate}
    \tau(x)=E\{Y(x)\}.
\end{equation}
We further use $\mu_x=E(Y|X=x,\bfZ)$ and $\pi_x=Prob(X=x|\bfZ)$ to denote the conditional mean (CM) of the outcome given exposure status $X=x$ and confounders $\bfZ$ and the conditional probability of $X=x$ given confounders $\bfZ$, respectively. In the literature, $\pi_x$ is often referred to as the \revisetwo{generalized} propensity score (PS) \citep{rosenbaum1983central}, \revisetwo{recognizing the multi-categorical nature of the exposure}. The fundamental problem in causal inference is that we only observe at most one potential outcome for each subject. Thus, to identify $\tau(x)$, we require the following two assumptions that are widely adopted {in the causal inference literature} \citep{rosenbaum1983central,yang2019combining}.

\textbf{Assumption 1} (no unmeasured confounding): $Y(x)\indep X|\bfZ$, \red{for $x=0,\ldots,L$}.

\textbf{Assumption 2} (positivity): $0<\pi_x<1$, \red{for $x=0,\ldots,L$}.

 \red{\revisetwo{In the next section}, we describe a \revisetwo{general} framework \revisetwo{that can be broadly applied to} estimate population mean potential outcomes in exposure groups, showing robust estimation with reduced variability compared to using the main data alone.}

 \begin{table}
\centering
\scriptsize{9}{11}\selectfont
\setlength{\tabcolsep}{0.5pt}
\caption{\revisethree{Comparison of UKB participants' characteristics in the primary analysis. Age was recorded at the first image visit, while other variables were measured at the baseline.}}
 \label{tab:demo}
\begin{tabular}{lllcc}
\toprule
\multicolumn{2}{l}{Continuous variables: Median (range)}     &      & The main data $(3,435)$ & The auxiliary data $(21,874)$ \\\midrule
Age, years                    &                                   &      & 64 (46-81)    & 65 (46-82)         \\
Body Mass Index, weight (kg)/$[$height (m)$]^2$                     &                                   &      & 26 (17-52)    & 26 (15-55)         \\ \toprule
\multicolumn{2}{l}{Categorical variables: Count $(\%)$}                &      & The main data $(3,435)$ & The auxiliary data $(21,874)$ \\ \midrule
Education               & \multicolumn{2}{l}{University or higher} & 1794 (52\%)   & 11352 (52\%)        \\
Income                  & \textless{}31000                  &      & 1031 (30\%)    & 7061 (32\%)        \\
Sex                  & Female                            &      & 1657 (48\%)   & 11700 (53\%)      \\\bottomrule
\end{tabular}
\end{table}

\section{Method}\label{Method}
\subsection{Augmented inverse probability of treatment weighting}\label{Augmented inverse probability weighting}
\revisetwo{To illustrate our proposed method, we start with a review of the well-known augmented inverse probability of treatment weighting (AIPTW) estimator: } 
\begin{equation}\label{aipw}
\begin{split}
        \hat{\tau}_{\text{aiptw}}(x)=\frac{1}{n}\sum_{i=1}^n\left\{\frac{I(X_i=x)}{\hat\pi_{xi}}Y_i-\frac{I(X_i=x)-\hat\pi_{xi}}{\hat\pi_{xi}}\hat\mu_{xi}\right\} \text{ for } \red{x=0,\ldots,L},
\end{split}
\end{equation}
where $I(E)$ is an indicator function that equals $1$ if event $E$ happens and $0$ otherwise. 
 Specifically, the term $\{I(X_i=x)/\hat\pi_{xi}\}Y_i$ corresponds to the estimated inverse probability of treatment weighting (\revisetwo{IPTW}) method accounting for imbalanced distribution of exposure assignment. 
 The term $[\{I(X_i=x)-\hat\pi_{xi}\}/\hat\pi_{xi}]\hat\mu_{xi}$ incorporates auxiliary information from covariates $\bfZ_i$ through the estimated CM $\hat\mu_{xi}$. 
A conventional approach to estimate the \red{PS} model and \red{CM} model is to specify parametric models $\pi_{x}(\bfalpha)$ and $\mu_x(\bfxi)$ with parameters $\bfalpha$ and $\bfxi$ solved by traditional regression methods \citep{nelder1972generalized}. \revisetwo{However, parametric models are too stringent when analyzing a large-scale and complex observational database and can lead to substantial bias if both models are mis-specified \citep{kang2007demystifying}.}

 \red{To \revisetwo{model more complex nonlinear mean structures in PS and CM while maintaining valid statistical inference}, recent research has incorporated machine-learning algorithms into the AIPTW framework, such as penalized regression and random forests, to fit richly parameterized models with flexible functional forms for the variables in $\bfZ_i$. The resulting estimator has potential to captures nonlinear confounding effects \revisetwo{while maintaining valid statistical inference \citep{hernan2010causal,chernozhukov2018double}.}} 
 \revisetwo{Despite these advancements,} however, \red{the machine-learning-based AIPTW approach itself cannot identify the best algorithm that leads to the most \revisetwo{reliable causal inference in practice}}, owing to the large number of available computational techniques. This challenge motivates ensembling multiple machine-learning algorithms, which we detail in the next subsection. 
We note that a more complex model does not necessarily yield a more accurate estimate of \revisetwo{mean potential outcomes}. For instance, when the true model indeed follows a linear structure, conventional regression may outperform more sophisticated models in terms of numerical accuracy. We refer readers to Section \ref{Simulation} for supporting numerical evidence.

\subsection{Robust weighting with machine learning}\label{Multiply robust weighting}

In this subsection, we propose a learner that allows multiple machine-learning algorithms to fit the \red{PS} and \red{CM} models and leads to valid statistical inference. This learner implicitly implements algorithm ensembling without requiring prior knowledge of the best algorithm. 

Before presenting the proposed learning procedure, we first introduce more notation. For $x=\red{0,\ldots,L}$, let $\hat{\pi}^{(1)}_{xi}, \ldots, \hat{\pi}^{(J_1)}_{xi}$ be the estimates of \red{PS} for subject $i$ based on a total of $J_1$ candidate algorithms. For example, $\hat{\pi}^{(1)}_{xi}$ is learned by \revisetwo{multinomial logistic} regression with $l_2$-penalty, $\hat{\pi}^{(2)}_{xi}$ is learned by random forest, and $\hat{\pi}^{(3)}_{xi}$ is learned by gradient boosting, etc. Similarly, let $\hat{\mu}_{xi}^{(1)}, \ldots, \hat{\mu}_{xi}^{(J_2)}$ be the estimates of \red{CM} based on $J_2$ candidate algorithms. Given all candidates, our robust causal machine learning (CML) estimator for \revisetwo{the exposure level $x$} is built upon the following weighted estimation: 
\begin{equation}\label{mr}
    \hat{\tau}_{\text{cml}}(x)=\sum_{i\in\{i|X_i=x\}}\hat{\omega}_{xi}Y_i, \text{ for } \red{x=0,\ldots,L},
\end{equation}
where the estimated weight $\hat{\omega}_{xi}$ is solved by maximizing $\prod_{i\in \{i|X_i=x\}}{\omega}_{xi}$ with respect to constraints ${\omega}_{xi}>0$, $\sum_{i\in \{i|X_i=x\}}{\omega}_{xi}=1$, and $\sum_{i\in \{i|X_i=x\}}{\omega}_{xi}\hat{\bfg}_{xi}=\bfzero$, with
\begin{equation}\label{g}
\begin{split}
    \hat{\bfg}_{xi}=\bigg(&\hat{\pi}^{(1)}_{xi}-\frac{1}{n}\sum_{l=1}^n\hat{\pi}^{(1)}_{xl}, \ldots \hat{\pi}^{(J_1)}_{xi}-\frac{1}{n}\sum_{l=1}^n\hat{\pi}^{(J_1)}_{xl},\\
    & \hat{\mu}^{(1)}_{xi}-\frac{1}{n}\sum_{l=1}^n\hat{\mu}^{(1)}_{xl}, \ldots \hat{\mu}^{(J_2)}_{xi}-\frac{1}{n}\sum_{l=1}^n\hat{\mu}^{(J_2)}_{xl}\bigg)^{\scriptscriptstyle \top}. 
\end{split}
\end{equation}
 In particular, the constraint $\sum_{i\in \{i|X_i=x\}}{\omega}_{xi}\hat{\bfg}_{xi}=\bfzero$ is imposed such that the weight $\hat{\omega}_{xi}$ \revisetwo{calibrates the covariate distribution} of the corresponding group to that of the entire population, 
 thus leading to a consistent estimator of the population mean potential outcome $\tau(x)$.
 \red{The original estimator} was first proposed by \cite{han2013estimation} \red{under the context of missing data}, in which generalized linear models were used to \red{fit missing data and outcome CM models}}. In this paper, we broaden its use \red{to causal inference with a multi-categorical exposure and} incorporate advanced machine learning into estimation.

\revisetwo{From a theoretical perspective}, \red{the estimator from \cite{han2013estimation}} has oracle convergence rate $O_p(n^{-1/2})$ only when conventional statistical regression is used to fit \red{PS} and \red{CM}. In the context of machine learning, this estimator may have a lower convergence rate, $O_p(n^{-\phi})$, with $\phi<1/2$ \citep{hernan2010causal,chernozhukov2018double}. The key driving force behind this behavior is the bias in learning the true $\pi_{xi}$ and $\mu_{xi}$. Similar to non-parametric estimation, machine learning estimation prevents the variance of the estimator from exploding at the price of induced bias. As such, we have $\hat{\pi}_{xi}^{(j_1)}-\pi_{xi}=O_p(n^{-\phi_1}),~j_1=1,\dots,J_1$ and $\hat{\mu}_{xi}^{(j_1)}-\mu_{xi}=O_p(n^{-\phi_2}),~j_2=1,\dots,J_2$, with $\phi_1,\phi_2<1/2$. Therefore, $\hat{\tau}_{\text{cml}}(x)$ may have a slower convergence rate and thus be less desirable for use in applications \citep{chernozhukov2018double}. 

\revisetwo{To achieve a desirable convergence rate, \red{establish a tractable asymptotic distribution} for $\hat{\tau}_{\text{cml}}(x)$ to guide statistical inference in practice, and \red{mitigate potential overfitting in PS and CM}}, we recommend employing sample splitting and cross-fitting (SSCF) for practical implementation.
For example, we can randomly split the data into a training set and an evaluation set, each with $n/2$ samples. We first apply predictive algorithms to samples in the training set to obtain the estimates $\hat{\pi}^{(j_1)}_{xi}$ and $\hat{\mu}^{(j_2)}_{xi}$ for $j_1=1,\ldots, J_1$ and $j_2=1,\ldots, J_2$, which are then evaluated on samples in the evaluation set to estimate $\hat{\tau}_{\text{cml}}(x)$. 
To fully make use of the data, we repeat the above procedure but swap the roles of the training and evaluation halves of the main sample; the final estimate will be the average of the two estimates for $\tau(x)$ from each half of the population. \red{The above process avoids the use of the same set of outcomes in both PS \& CM and mean potential outcome estimation, enabling valid statistical inference \citep{hernan2010causal,chernozhukov2018double}.} 



\subsection{Information integration }\label{Information integration}
The proposed estimator, $\hat{\tau}_{\text{cml}}(x)$,  in (\ref{mr}) is robust and has little bias under mild conditions. We refer readers to Section \ref{Theoretical property} and Section 1.2 of the Supplementary Material for details. Thus, it successfully achieves our first goal, i.e., robustness. \revisetwo{Motivated by our case study, we now explore the potential to} improve efficiency by integrating information from auxiliary data on participants $i=n+1,\ldots,N$, while maintaining robustness. We consider cases where the $p-$dimensional confounder vector $\bfZ$ is not observed at all or only partially observed in the auxiliary data. Consequently, naively merging the main and auxiliary datasets and using only the observed variables would lead to a biased estimate of the mean potential outcome. Given the complex nature of the proposed CML estimator, integrating information from the auxiliary data requires an alternative approach. To this end, we propose the following causal machine learning estimator with information borrowing (CMLIB),
\begin{equation}\label{mrib}
    \hat{\tau}_{\text{cmlib}}(x)=\sum_{i\in\{i|X_i=x\}}\frac{\hat{p}_i\hat{\omega}^\ast_{xi}Y_i}{\sum_{i\in\{i|X_i=x\}}\hat{p}_i\hat{\omega}^\ast_{xi}}.
\end{equation}
\red{In contrast to the estimator in (\ref{mr}), CMLIB incorporates two weights. One is $\hat{p}_i$ (named ``integration scores"), and the other is the calibration weight $\hat{\omega}^\ast_{xi}$. The rationale of proposed estimator (\ref{mrib}) and its corresponding estimation procedure are summarized below.

\textbf{Integration scores.} Integration score is a new concept developed in this paper to deliver information from auxiliary data. Using both main and auxiliary data, the integration scores $\hat{p}_{i}$ are estimated by maximizing $\prod_{i=1}^N p_i$ with respect to 
$ p_i>0$, $\sum_{i=1}^N{p}_i=N$, and $\sum_{i=1}^N{p}_i\bfH_i(\bftheta)=\bfzero$ with
\begin{equation}\label{h}
    \bfH_i(\bftheta)=\begin{Bmatrix}
R_i{\bfh}_{i}(\bftheta)\\
(1-R_i){\bfh}_{i}(\bftheta)
\end{Bmatrix}.
\end{equation}
Here, $R_i$ is a data indicator that takes the value of $1$ if $i=1,\ldots,n$ (main data) and 
$0$ if $i=n+1,\ldots,N$ (auxiliary data). The function ${\bfh}_{i}(\bftheta)$ is a ``working" estimating function with a nuisance parameter vector $\bftheta$ for both main and auxiliary data. Two remarks are highlighted: (1) the working function ${\bfh}_{i}(\bftheta)$ should be constructed only based on the observed variables from both data, and (2) different functional forms of ${\bfh}_{i}(\bftheta)$ may affect integration performance. Herein, we introduce two simple forms (but not limited to these two in practice) that work well in our simulation and real data applications. 
\begin{equation}\label{htilde}
  (I)~ {h}_{i}(\theta)=Y_i-\theta ~\text{and} ~ (II)~{\bfh}_{i}(\bftheta)=(1, X_i)^{\scriptscriptstyle \top}\{Y_i-(1, X_i)\bftheta\}.  
\end{equation}
The latter function is based upon the notation with a binary exposure. The function in the setting with a multi-categorical exposure can be similarly constructed. 

The intuition of proposing integration scores relies on the construction of the over-identified estimating function in (\ref{h}). Here, over-identification means that the length of indexed parameter vector $\bftheta$ should be shorter than the length of the estimating function $\bfH_i(\bftheta)$. Motivated by empirical likelihood theory \citep{qin1994empirical,chen2023efficient}, an over-identified function will enable the resulting estimates $\hat p_i$ to encapsulate information from the used variables from both datasets, which produces a more efficient estimate of the distribution function of these variables than the simple empirical distribution function \citep{qin1994empirical}. As a result, using estimated scores $\hat p_i$ as integration weights in (\ref{mrib}) (compared to weights equal to $1$) is expected to facilitate the transfer of information to the estimation of mean potential outcomes.

We also remark here that the shared function $\bfh_i(\cdot)$ \revisetwo{and a consistent $\bftheta$} in both datasets can be achieved if the main and auxiliary data are from the same population. Violating this condition may lead to biased estimation for population mean potential outcomes. In real cases where baseline and observed covariate distributions are not the same in the main and auxiliary data, pre-screening the auxiliary data is recommended to address inclusion and exclusion criteria in the main study. \revisetwo{We also recommend conducting exploratory data analysis and a hypothesis test to assess if the parameters $\bftheta$  are equal between the main data and the auxiliary data. A small $P$-value may indicate heterogeneity between the two datasets.} Alternatively, we may implement PS matching based on observed covariates to correct imbalanced distributions between the two datasets so that the two datasets can be more similar to each other. More discussions of handling heterogeneous populations can be found in Section \ref{Evaluation} and \revisetwo{Sections 3.2.2 and 4 of the Supplementary Material}. 

 \textbf{Calibration weights.} These weights aim to balance covariate distributions in a robust manner. One naive attempt is to directly use the weights $\hat{\omega}_{xi}$ in (\ref{mr}). However, doing so does not ensure obtaining theoretical variance reduction for the estimate in (\ref{mrib}). To achieve efficiency gain, we consider the following modification: after obtaining integration scores $\hat p_i$, we can calculate the calibration weights $\hat{\omega}^\ast_{xi}$ by maximizing $\prod_{i\in \{i|X_i=x\}}{\omega}^\ast_{xi}$ with respect to ${\omega}^\ast_{xi}>0$, $\sum_{i\in \{i|X_i=x\}}{\omega}^\ast_{xi}=1$, and $\sum_{i\in \{i|X_i=x\}}{\omega}^\ast_{xi}\hat{\bfg}^\ast_{xi}=\bfzero$, where
\begin{equation}\label{g ast}
\begin{split}
    \hat{\bfg}^\ast_{xi}=\bigg(&\hat{\pi}^{(1)}_{xi}-\frac{1}{n}\sum_{l=1}^n\hat{\pi}^{(1)}_{xl}, \ldots, \hat{\pi}^{(J_1)}_{xi}-\frac{1}{n}\sum_{l=1}^n\hat{\pi}^{(J_1)}_{xl},
    \hat{p}_i\hat{\mu}^{(1)}_{xi}-\frac{1}{n}\sum_{l=1}^n\hat{p}_l\hat{\mu}^{(1)}_{xl} +(1-\hat{p}_i)\hat{\eta}_x^{(1)},\\
    & \ldots, \hat{p}_i\hat{\mu}^{(J_2)}_{xi}-\frac{1}{n}\sum_{l=1}^n\hat{p}_l\hat{\mu}^{(J_2)}_{xl}+(1-\hat{p}_i)\hat{\eta}_x^{(J_2)}\bigg)^{\scriptscriptstyle \top}
\end{split}
\end{equation}
 and $\hat{\eta}_x^{(j_2)}=(1/n)\sum_{l=1}^n\hat{\mu}^{(j_2)}_{xl}$, for $j_2=1,\ldots,J_2$. The updated calibration function $\hat{\bfg}^\ast_{xi}$ in (\ref{g ast}) incorporates integration scores $\hat{p}_{i}$ and $(1-\hat{p}_i)\hat{\eta}_x^{(j_2)}$, contrasting with the calibration function $\hat{\bfg}_{xi}$ in (\ref{g}). This modification is rooted in mathematical reasoning aimed at ensuring reduced estimation variability. We refer readers to technical details in Section 1.3 of the Supplementary Material. 

We remark here that the doubly weighted estimator in (\ref{mrib}) serves as a generalization of the CML estimator in (\ref{mr}). Specifically, in scenarios without auxiliary data (i.e., when $N=n$), the integration scores will be reduced to $1$ as per empirical likelihood theory \citep{qin1994empirical}. Consequently, the calibration weights $\hat{\omega}^\ast_{xi}$ in (\ref{mrib}) will be reduced to $\hat{\omega}_{xi}$ in (\ref{mr}), i.e., no efficiency gain at all.  Also, it is noteworthy that SSCF is unnecessary for calculating integration scores, but they are required for computing calibration weights. The complete learning procedure is summarized in Algorithm \ref{estimation}. 
\begin{algorithm}
\SetAlgoLined
\caption{{Estimation Procedure}}\label{estimation}
\scriptsize
\red{
\textbf{Data}: Main data $(Y_i,X_i,\bfZ_i)_{i=1}^n$ and Auxiliary data $(Y_i,X_i)_{i=n+1}^N$\;
\uIf{Auxiliary data are not available}{
    Set $\hat{p}_i\leftarrow 1$, for $i=1,\ldots,n$ \;
  }
  \Else{
  \textbf{Integration scores}: Obtain the scores $\hat{p}_i$ by (\ref{h}) based on both main data and auxiliary data \;
  \textbf{Data}: Randomly split the main data into two halves, denoted by data $1$ and data $2$\;
    \For{each data $i\in\{1,2\}$}{
Apply different candidate algorithms to learn $\hat{\pi}_{x}$ and $\hat{\mu}_{x}$ based on data $i$ \;
Calculate the estimator $\hat{\tau}_{\text{cmlib}}(x;j)$ based on (\ref{mrib}) and the data $j$, $j\neq i$\;
}
  }
\textbf{Result}: Obtain the final estimator $\hat{\tau}_{\text{cmlib}}(x)=\{\hat{\tau}_{\text{cmlib}}(x;1)+\hat{\tau}_{\text{cmlib}}(x;2)\}/2$.
}
\end{algorithm}


 }


\revisetwo{\subsection{Theoretical properties and their practical implications}\label{Theoretical property}
In this subsection, we provide a high-level summary of key properties that enhance the practical applicability of the proposed learner (CML and CMLIB). All mathematical details can be found in the Supplementary Material.

\textbf{Robust weighting for causal inference}. Unlike existing ensemble learning approaches, such as Super Learner \citep{van2007super}, which integrate multiple algorithms to optimize outcome prediction, 
 the proposed weighting methods in (\ref{mr}) and (\ref{mrib}) are specifically designed to integrate multiple algorithms for calibrating imbalanced covariate distributions resulting from missing data \citep{han2013estimation} or non-random exposure assignment. Moreover, the proposed weights are robust to near-zero PS values due to the incorporation of nonnegative and sum-to-one constraints, as well as the implicit involvement of additional smoothing weights (described in Section 1.3 of the Supplementary Material) that further mitigate the impact of large inverse propensity score values on numerical performance \citep{han2013estimation,han2014multiply}. Consequently, the proposed estimators are expected to demonstrate greater robustness to extreme weights compared to the IPTW and AIPTW approaches.

\textbf{Evaluation of covariate balance}. 
A key advantage of the proposed calibrated weights ($\hat{\omega}_{xi}$ and $\hat{\omega}_{xi}^\ast$) is their ability to balance covariate distributions across exposure groups (Section 1.3 of the Supplementary Material). This property enables validation of the weighting scheme by assessing the balance of baseline confounders between groups: improved covariate balance supports the validity of downstream causal inferences using these weights.  
We adopt a standard approach from PS matching or weighting approaches. Specifically, we calculate the {standardized mean difference} (SMD) for each covariate using the weighted samples. Following established guidelines \citep{zhang2019balance}, an SMD value below 0.1 indicates adequate balance.

\textbf{Statistical inference}. 
In addition to the two assumptions above, we require  three additional conditions to facilitate statistical inference for the CML and CMLIB estimators in practical settings. First, the PS and CM estimators are ``well-behaved" to achieve the desired convergence rates. Second, the estimation framework assumes a ``robust learning guarantee", meaning at least two candidate algorithms within the ensemble can consistently approximate the true PS and CM. Third, for CMLIB specifically, the main and auxiliary populations must exhibit homogeneity in covariate distributions to preclude selection bias. Under these assumptions, we establish that both estimators are asymptotically normal, with the variance of the CMLIB estimator no larger than that of the CML estimator (Section 1.2 of the Supplementary Material). To estimate standard errors and enable robust statistical inference, we recommend a non-parametric bootstrap procedure. 
To ensure independent validation, all tuning parameters are fixed during estimation based on bootstrapped data. The pseudo-code for this procedure is summarized in Algorithm S1 in the Supplementary Material. We have also evaluated additional theoretical properties, including an alternative efficient algorithm for estimating the nuisance parameters $\bftheta$, the optimal utilization of auxiliary data, and a method extension in which only summary statistics are available in the auxiliary data (Sections 1.2, 1.4, 1.5, and 4 of the Supplementary Material). 
}

\section{Alcohol intake frequency and \revisethree{brain white matter microstructural integrity} in the UKB }\label{A real data application}
\cite{mende2019alcohol} summarized the effect of alcohol intake on cognitive decline and noted that heavy drinkers may experience greater average cognitive decline than mild-to-moderate drinkers. \red{In a meta-analysis, \cite{zhao2023association} reported that low or moderate drinkers showed no significant association with all-cause mortality risk compared to lifetime nondrinkers, whereas an increased risk was observed at higher intake levels. In our study, we further investigated the impact of alcohol intake frequency on brain white matter microstructure integrity among middle aged and older adults enrolled in the UKB study}. Instead of cognition \red{or mortality} outcomes, which are consequences of brain aging and white matter degredation, \revisethree{we focused on a more {objectively measured} neuroimaging biomarker: fornix FA. The fornix shows substantial microstructure degradation with aging \citep{chen2015age}, and reductions in fornix FA have been associated with poorer memory performance and {neurodegenerative disorders} \citep{kantarci2014fractional, bangen2021decreased}. Our primary objective was to estimate the mean potential outcome of fornix FA and compare these values across different levels of alcohol intake frequency.}

\textbf{Validity of the primary outcome as a brain aging marker.} \revisethree{The fornix FA values from white matter fiber tracts were extracted using diffusion MRI, where higher values indicate better microstructural integrity}. For more data processing details, we refer readers to Section~2.1 of the Supplementary Material. \revisethree{In addition to evidence from the literature, we conducted additional assessments of the association between fornix FA and aging-associated characteristics using UKB,} with results summarized in Figure~\ref{fig:workflow}. \revisethree{We observed the strongest negative correlation between chronological age at the first image visit and fornix FA among all $39$ white matter tracks, with a Kendall-tau correlation coefficient of $-0.375$, which was significantly below zero (P-value$<0.001$, Figure~\ref{fig:workflow}A and Figure~S.1 of the Supplementary Material)}. This finding suggests that fornix FA may serve as an informative marker of brain aging.  \revisethree{Fornix FA} also demonstrated robust associations with multiple cognitive performance measures: \revisethree{reduced Fornix FA} was tied to fewer digits matched (M1), lower fluid intelligence scores (M2), fewer maximum digits matched (M3), fewer puzzles solved correctly (M4), more incorrect matches in a round (M5), and longer times to complete matching tasks (M6) and find correct paths (M7) (Figure~\ref{fig:workflow}B), \revisethree{with some Kendall’s tau values approaching $\pm 0.2$}. Together, these findings provide extra quantitative evidence for using fornix FA as a meaningful marker of brain aging-related outcomes, highlighting its potential utility in neuroimaging research \citep{chen2015age, bangen2021decreased}. Additionally, we observed that \revisethree{lower fornix FA} was significantly associated with elevated systolic blood pressure (SBP$>140$ mmHg) (Figure~\ref{fig:workflow}C)  and diastolic blood pressure (DBP$>90$ mmHg) (Figure~S.2).



\textbf{Secondary outcomes.} \revisetwo{ In addition to the primary outcome, we also analyzed several aging-associated variables as secondary outcomes, including measurements of SBP (Figure~\ref{fig:workflow}C) and DBP (Figure S.2) \citep{pettersen2014arterial}, as well as two cognitive performance metrics, M1 and M7, that exhibited the highest association with fornix FA (Figure~\ref{fig:workflow}B).
}

\textbf{Alcohol intake and phenomic variables.} \revisetwo{The data on alcohol intake frequency were collected through a UKB-designed survey using a touchscreen questionnaire. The questionnaire included seven response options: 1) almost daily, 2) three to four times per week, 3) once or twice a week, 4) one to three times a month, 5) special occasions only, 6) never, and 7) prefer not to answer. To ensure sufficient sample sizes in each group and clinically meaningful categories, we consolidated these responses into three exclusive subgroups: ``heavy" drinkers (1), ``moderate" drinkers (2, 3), and ``light/never" drinkers (4, 5, 6). 


}

We considered phenomic variables as potential confounders to systematically account for the social, health, and physical conditions of participants. These included basic profiles (\revisethree{chronological age at the first image visit}, sex, education, household income, body mass index), nutrition intake (cooked/raw vegetables and/or fruits), blood biochemistry ($25$ measures), physical activity/measures ($4$ measures), nuclear-magnetic-resonance metabolomics ($168$ measures), and urine assays ($3$ measures). \revisethree{All phenomic variables were measured at baseline (instance 0, $2006$-$2010$), which occurred on average $8$ years prior to the first imaging visit (instance 2, $2014$-$2018$), to mitigate the risk of reverse causality}. Baseline cognition metric (M6) was also included to account for initial brain conditions. Simulation results (Tables~S.1-S.2, S.4-S.7) demonstrated that including too many noisy covariates may increase estimation bias. High missingness in blood biochemistry, metabolomics, and physical measures could further reduce the sample size if all variables were included. Thus, we implemented a pre-screening step to identify more plausible confounders with significant Kendall-tau correlation with both the primary outcome and the exposure. After screening, \revisethree{$89$ baseline confounders remained. Owing to missingness in the remaining confounders, the final main dataset included $3,435$ participants: $860$ heavy drinkers, $1,975$ moderate drinkers, and $600$ light/never drinkers}.
Additionally, \revisethree{$21,874$} participants with missing or incomplete phenomic data were treated as auxiliary data. \revisetwo{A summary of the study population is presented in Table \ref{tab:demo}, where we observed similar basic profiles between the main and auxiliary data.}
\revisethree{We computed estimates using our proposed methods, CML and CMLIB, as well as eight competing methods: simple average, single algorithm-based AIPTW using penalized regression, random forest, gradient boosting, respectively, and Super Learner-assisted outcome regression, IPTW, AIPTW, and TMLE}, for analyzing the primary outcome. CMLIB utilized the $\bfh$ function $(II)$ in (\ref{htilde}) to estimate mean potential outcomes for each exposure level. \red{The same learning and tuning procedures were applied for all methods, as detailed in Section \ref{Competing Methods}.} \revisethree{The $95\%$ Confidence intervals were calculated using quantiles from $400$ bootstrap samples.}

\red{\textbf{Validity of assumptions.} \revisetwo{To assess the validity of our assumptions, we first evaluated covariate balance using SMD and found that the proposed weighting strategies substantially improved balance, with SMD $< 0.1$ for most covariates (Figure~\ref{fig:bag}A and Figure~S.3). This finding is consistent with a valid causal interpretation of our estimates, assuming no unmeasured confounders. Next, we examined the distribution of generalized PS and observed that they were well distributed and remained above $0.01$ for nearly all subjects (Figure~\ref{fig:bag}B), supporting the validity of the positivity assumption. Finally, we conducted a Wilcoxon rank-sum test to compare the $\bftheta$ estimates between the main and auxiliary data and found no significant differences (\revisethree{$P = 0.992$, $0.193$, and $0.687$} for the outcome means of the light/never, moderate, and heavy groups, respectively), which justifies the use of auxiliary data. Collectively, these results are consistent with our assumptions.}
}

\begin{figure}
    \centering
    \includegraphics[width=5in]{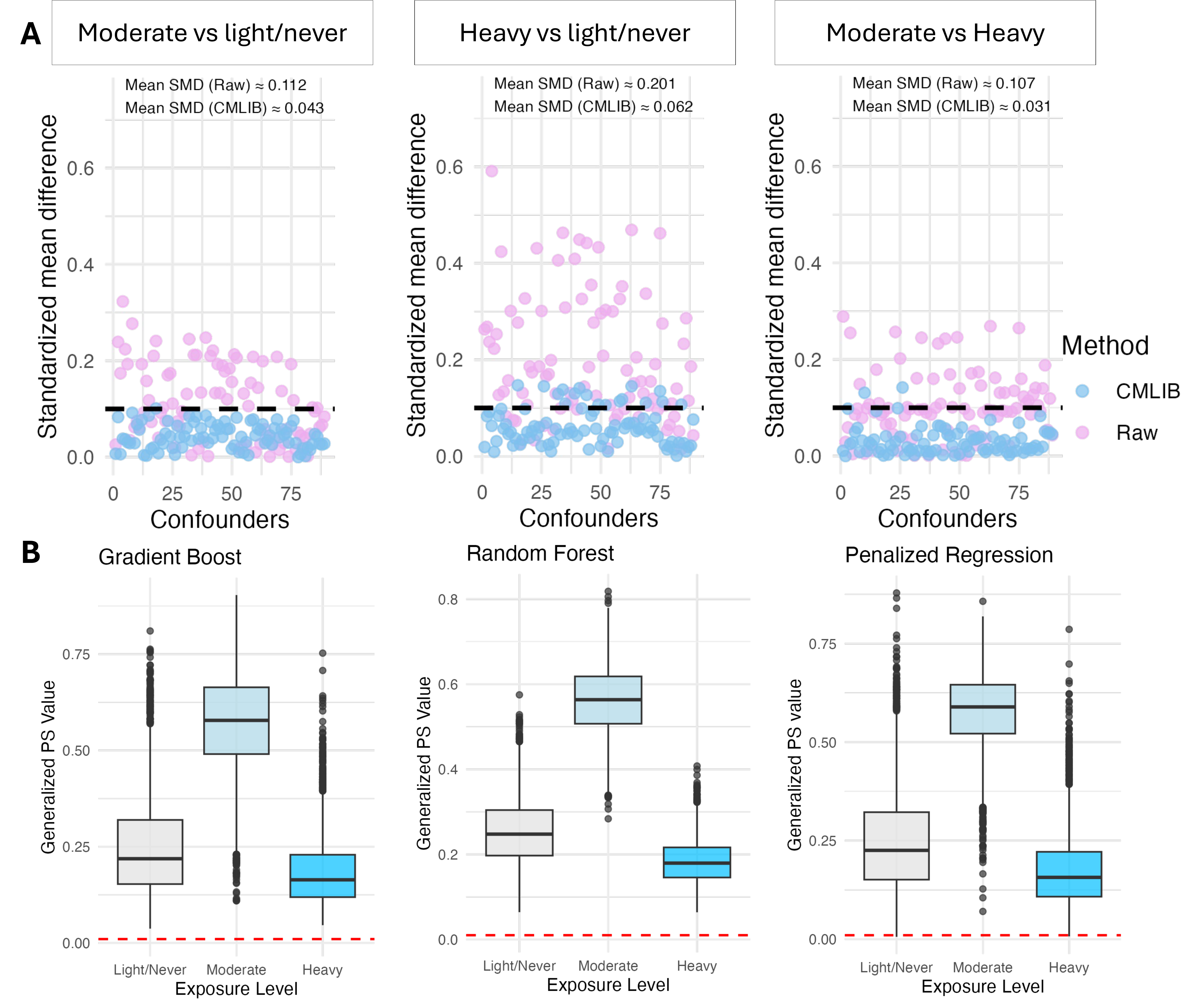}
    \caption{A. Assessment of the standardized mean differences in covariates between any two of the three groups. The black dashed line represents SMD $= 0.1$. B. Assessment of the positivity assumption based on three machine learning approaches.
The red dashed line represents $0.01$ generalized PS values. }
    \label{fig:bag}
\end{figure}

\red{
\textbf{Sensitivity analysis for the primary outcome.} To evaluate the rigor of our findings, we conducted four sensitivity analyses. \revisetwo{First, we compared the results with matching the main and auxiliary data using a matching ratio of 1:5. Second, we evaluated the sensitivity of the proposed confidence intervals by comparing them with those constructed using bootstrap standard errors and normal distribution quantiles. Third, we assessed the impact of the proposed covariate pre-screening by comparing the results with those obtained using all potential confounders without screening}. \revisethree{Lastly, we repeated the analysis excluding the baseline cognition metric from the confounder pool to evaluate the potential over-adjustment.} We refer interested readers to Section~2.4 of the Supplementary Material for detailed information on the sensitivity analysis.
}


\red{\textbf{Results.} The estimated mean potential fornix FA for ``heavy", ``moderate", and ``light/never" alcohol intake frequencies is summarized in Figure~\ref{fig:fornix fa}. All estimators showed little evidence that ``moderate" alcohol intake causes lower estimated mean potential fornix FA than ``light/never" alcohol intake. However, \revisethree{nine out of ten estimators indicated that ``heavy" alcohol intake caused significantly lower mean potential fornix FA than ``moderate" or ``light/never" alcohol intake. Notably, the magnitude of the estimated difference for mean potential outcomes in fornix FA between the heavy drinking group and the other two groups was substantially smaller in magnitude for estimators that adjusted for phonemics data. This attenuation likely reflects the confounding effects of baseline factors, such as age, lifestyle behaviors, and health-related conditions.}

\revisethree{For the SBP secondary outcome (Figure~S.4), eight out of ten estimators identified significantly higher mean potential SBP in the ``heavy" drinking groups compared to the ``light/never" group, and six estimators detected a significantly higher mean potential SBP in the ``moderate" drinking group relative to the ``light/never" group. For DBP (Figure~S.5), all ten estimators revealed significantly higher mean potential DBP in both the ``heavy" and ``moderate" drinking groups compared to the ``light/never" group.} However, no significant differences in mean potential cognition metrics were observed among the exposure groups (Figure~S.6-S.7), which may be attributed to the predominantly healthy status of most UKB participants at the time of brain imaging. Notably, neurodegenerative conditions often have a long preclinical period, during which early deficits are difficult to detect using common cognitive measures.

\revisethree{Furthermore, the sensitivity analyses demonstrated the robustness of the primary analysis: analyses with PS matching (Figure~S.8), confidence intervals using quantiles from the standard normal distribution (Figure~S.9), the inclusion of all $210$ covariates without filtering (Figure~S.10), and exclusion of the baseline cognition metric from the confounder pool (Figure~S.11) all yielded similar estimates and significant levels. Collectively, these results consistently demonstrated a pattern of ``heavy" alcohol intake causing lower mean potential fornix FA as compared with ``light/never" alcohol intake.

\begin{figure}
    \centering
    \includegraphics[width=5.5in]{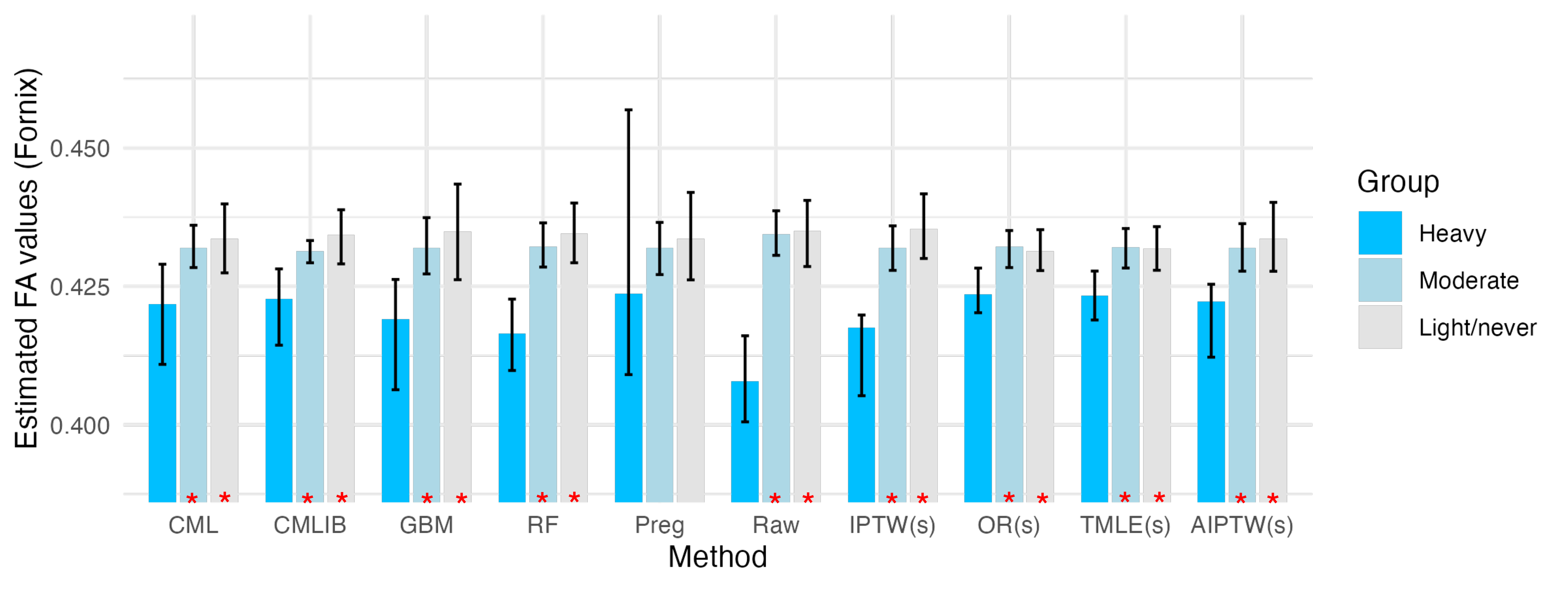}
    \caption{Mean potential outcome estimates of fornix FA for three alcohol intake frequencies. CML and CMLIB: the proposed estimator without and with information borrowing, respectively; Raw: simple average; AIPTW-based estimates with random forest (RF), $l_2$ penalized regression (Preg), or gradient boosting (GB);  Super Learner-assisted outcome regression IPTW, AIPTW, and TMLE
(OR(s), IPTW(s), AIPTW(s), TMLE(s)). $*$ indicates significant difference in fornix FA compared to the heavy alcohol intake group, based on a $95\%$ confidence interval using bootstrap quantiles.}
    \label{fig:fornix fa}
\end{figure}

Moreover, in most analyses, the CMLIB estimator exhibited much smaller bootstrapped standard deviation compared to CML without data integration. Also, both the CML and CMLIB estimators yielded very similar bootstrap quantile-based confidence intervals and normal approximation-based bootstrap confidence intervals, suggesting that the resulting estimators were approximately normally distributed. These findings support the validity of Theorems 1.1 and 1.2 in the Supplementary Material.}

}

\section{Simulation}\label{Simulation}

\subsection{Data generation}\label{Data generation}
\revisetwo{To closely represent the settings of our real data, we examined a scenario involving a $3$-level exposure ($X = 0, 1, 2$, representing ``light/never", ``moderate", and ``heavy" alcohol intake, respectively) with $100$ candidate confounders, among which $15$ are true confounders. For subject $i = 1, \ldots, n$ in the main data, we first generated confounders $\bfZ$ from a $p$-variate normal distribution with a mean of $0$, variance of $1$, and correlation coefficients of $0.5$ between each pair of confounders. Exposure $X$ was generated using a multinomial distribution with success probability $\pi_x = P(X = x \mid \bfZ)$ and marginal probabilities of $0.24$, $0.5$, and $0.26$, respectively. These probabilities were selected to reflect similar characteristics observed in the UKB. The potential outcome $Y(x)$ was then generated from a normal distribution with a conditional mean $\mu_x = E(Y(x) \mid \bfZ)$, and the variance was tailored to be $9.6$ and $36$ in order to achieve signal-to-noise ratios of $0.8$ and $0.2$, representing high and low outcome signals, respectively.  Additionally, we considered two cases in which the true PS and CM have different mean structures: both linear (case 1) and both non-linear (case 2), with mixture effect sizes for different confounders. These two cases reflect complexities of PS and CM that may exist in real data. For detailed formulas, we refer readers to Section 3.1 of the Supplementary Material. For subject $i=n+1,\ldots,N$ in the auxiliary data, we adopted the same data generating mechanism, except that we made $\bfZ_i$ a vector of unobserved confounders in the auxiliary data. 
The above data generation process guarantees a homogeneous population between main and auxiliary datasets. 
In both cases, we conducted $500$ Monte Carlo replicates. For each replicate, we considered sample size \revisethree{$n=3,000$} for the main data and sample size $5n$ for the auxiliary data.
}

\subsection{Competing methods and evaluation}\label{Competing Methods}
We implemented two proposed estimators: the CML estimator described in Section \ref{Multiply robust weighting} without information integration and the CMLIB estimator proposed in Section \ref{Information integration} with auxiliary data integration. The performance of our proposed estimators was compared with three competing methods that use the AIPTW estimator with different algorithms for estimating nuisance parameters. The first method fitted $l_2$-penalized logistic regression to the PS model and $l_2$-penalized linear regression to the CM model (AIPTW.Preg), with tuning parameters selected via five-fold cross-validation. The second method used random forest, averaging over $1,000$ trees, to estimate both the PS and CM models (AIPTW.RF). The third method applied gradient boosting to estimate the PS and CM models, with regularization parameters selected via ten-fold cross-validation (AIPTW.GB). Our CML and CMLIB estimators were implemented using all three algorithms ($l_2$-penalized regression, random forest, and gradient boosting). For the CMLIB estimator, we employed form $(I)$ in (\ref{htilde}) to facilitate information integration from auxiliary data. \revisethree{We additionally evaluated the performance of other existing approaches, specifically, outcome regression, IPTW, TMLE, and AIPTW, enhanced by the SuperLearner algorithm \citep{van2007super}, which ensembles the previous mentioned learners ($l_2$-penalized regression, random forest, and gradient boosting).} These ensemble-based estimators are denoted by OR(s), IPTW(s), TMLE(s), and AIPTW(s). We assessed the performance of all estimators based on estimation bias, Monte Carlo variability, MSE, and the distributions of the estimators.

\subsection{Results}\label{Evaluation}
\revisetwo{Estimation bias and variability are summarized in Figures \ref{fig:case1high} (Case 1) and \ref{fig:case3high} (Case 2). Regarding estimation bias, we observed that among the AIPTW-based estimators using a single algorithm, AIPTW.Preg performed the best when the PS and CM followed a linear structure, while AIPTW.RF slightly  outperformed AIPTW.Preg for estimates under exposure levels $0$ and $1$ when the PS and CM followed a non-linear structure. \revisethree{These results suggest that machine learning models exhibit context-dependent performance and may not always yield superior results across all scenarios}. Across all scenarios, regardless of the mean structures for PS and CM or the signal-to-noise ratio, CML and CMLIB consistently demonstrated the smallest or near-smallest estimation bias among the nine estimators. This result underscores the robustness of the proposed causal machine learners, which effectively integrate multiple algorithms to facilitate causal inference. In contrast, the four estimators utilizing the SuperLearner algorithm showed varied performance. Compared to CML and CMLIB, most of these methods exhibited larger bias 
when estimating the mean potential outcome of exposure level $0$ and $1$, particularly in case 1. \revisethree{Among the four SuperLearner-based estimators, IPTW(s) showed the largest bias, while AIPTW(s) demonstrated the smallest bias in most scenarios and performed consistently across exposure levels $1$ and $2$. OR(s) and TMLE(s) showed mild to moderate bias in most scenarios.} Notably, CML and CMLIB estimators yielded even smaller bias than the AIPTW(s) estimator in many scenarios in case 1 and performed comparably in case 2. This may be partially attributed to the greater appropriateness of using the calibration technique over predictive modeling for algorithm ensembling in causal inference, as well as the robustness of CML-based estimators against near-zero PS values \citep{han2013estimation}. 

\begin{figure}
    \centering
    \includegraphics[width=5in]{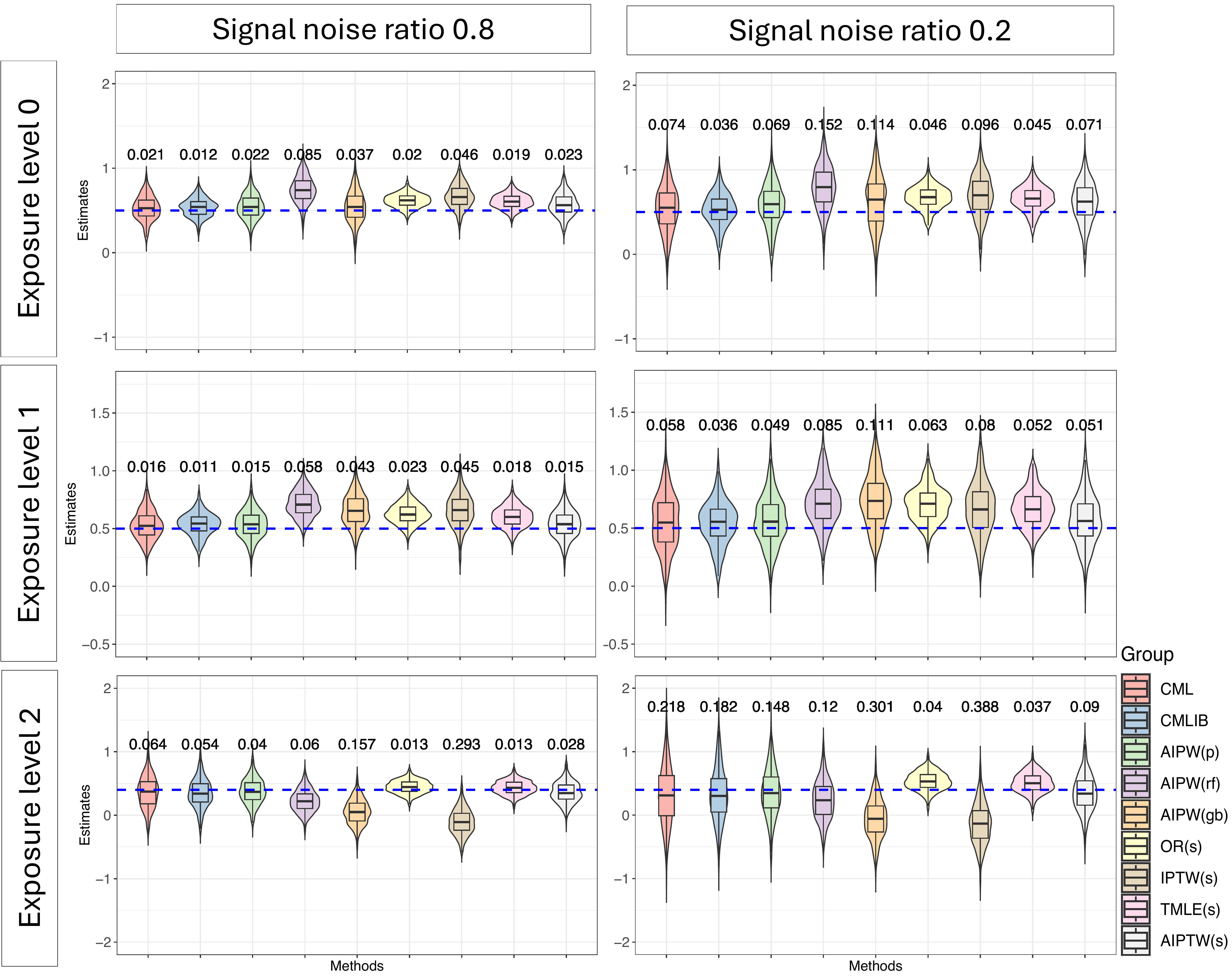}
    \caption{\revisethree{Evaluation of estimation bias based on Case 1 (linear) and $500$ Monte Carlo runs. MSEs are reported in each plot.  The blue dashed line represents the true value.}}
    \label{fig:case1high}
\end{figure}

Regarding estimation variability, we observed that the CMLIB estimator demonstrated lower variability than the CML estimator after incorporating auxiliary data, aligning with our theoretical findings in Section \ref{Theoretical property} and underscoring the benefits of integrating auxiliary data. \revisethree{It also achieved the smallest MSE among all estimators in several scenarios.}  Additionally, both CML and CMLIB estimators exhibited increased variability as the signal-to-noise ratio decreased. Among remaining estimators, \revisethree{IPTW(s) showed the largest MSE across all settings, while OR(s) consistently demonstrated the smallest variances. Notably, 
the OR(s) and TMLE(s) achieved comparable MSEs across most settings and exposure levels. This similarity may be attributed to noisy ensemble-based PS estimates in our simulation settings, which likely resulted in near-zero fluctuation parameters during the TMLE(s) targeting step, leading to TMLE(s) that were numerically close to those from OR(s).}

In summary, the CMLIB estimator is more desirable for minimizing estimation bias with decent variability, whereas the OR(s)-based estimator may be preferred for achieving the smallest variance, albeit with some degree of bias. Practical guidance for using these two estimators is provided in Section \ref{Discussion}.}

\begin{figure}
    \centering
    \includegraphics[width=5in]{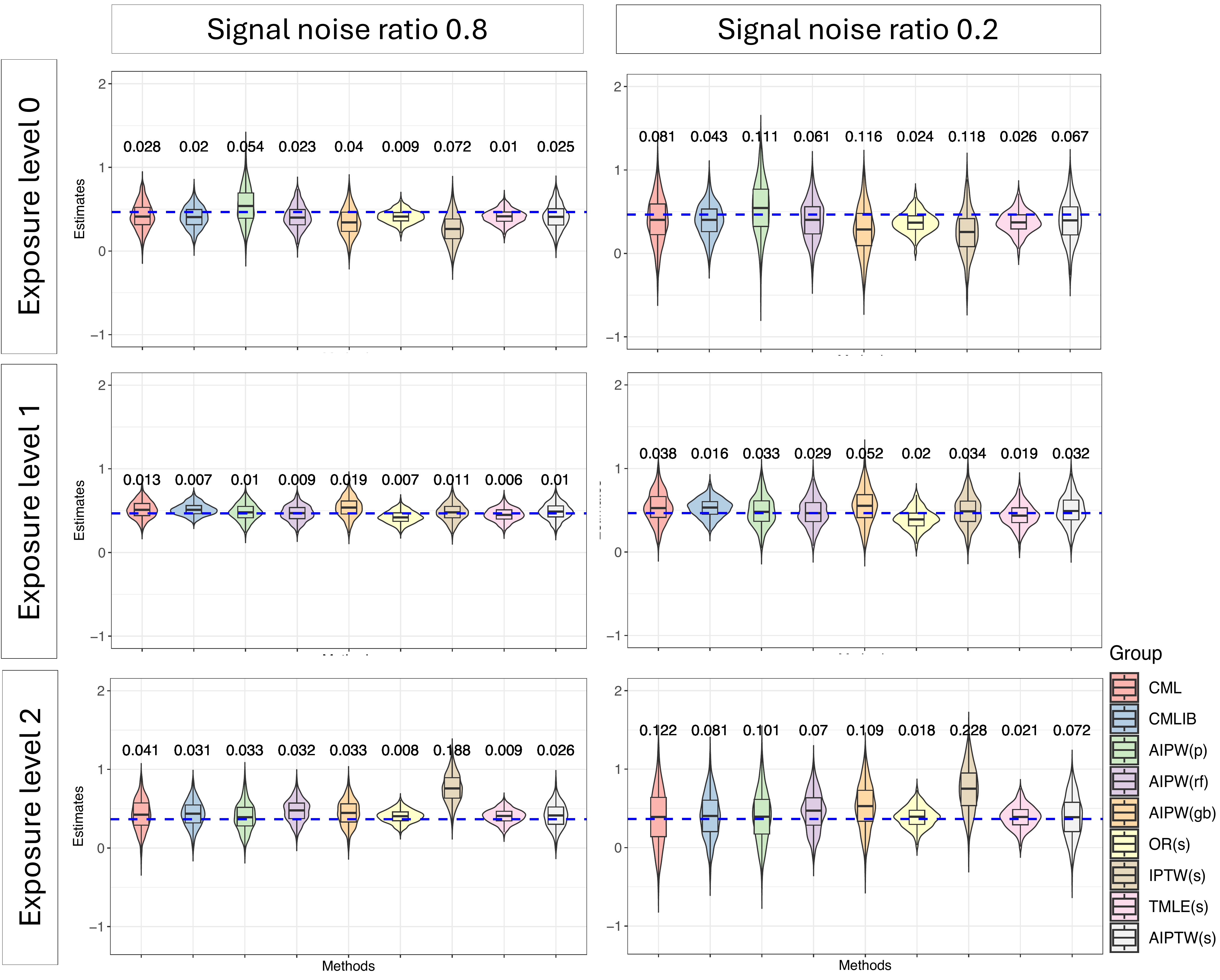}
    \caption{\revisethree{Evaluation of estimation bias based on Case 2 (nonlinear) and $500$ Monte Carlo runs. MSEs are reported in each plot.  The blue dashed line represents the true value.}}
    \label{fig:case3high}
\end{figure}

}

\revisetwo{In addition to the data generation scheme described above, we evaluated the performance of CML and CMLIB under other data settings, including a two-level exposure (Tables~S.1-S.2, S.4-S.7, S.9). We also explored alternative working functions (Table~S.10) and auxiliary data with a distribution different from that of the main data (Table~S.8). The first two settings led to the same conclusion: CML and CMLIB exhibited the lowest bias among single algorithm-based AIPTW estimators, with CMLIB being more efficient than CML after integrating additional information. Additionally, the bootstrap standard errors for CML and CMLIB were close to their empirical standard errors, and most of the constructed confidence intervals showed desirable coverage proportions (Table~S.3). In the last setting, we observed a substantial reduction in estimation bias after PS matching. We refer readers to Section~3.2 of the Supplementary Material for more details.

 }




\section{Discussion}\label{Discussion}
 This work introduces versatile causal learning methods that implicitly and effectively performs algorithm ensembling for causal inference, providing consistent and robust estimation of mean potential outcomes with tractable statistical inference. Additionally, the methods facilitate integration of auxiliary data, accommodating scenarios where large sets of phenomic variables may be completely missing or only partially observed, a common challenge in real-world studies. \revisetwo{Beyond continuous outcomes, the proposed estimation methods also accommodate other outcome types, including binary outcomes, while ensuring estimates remain bounded within the outcome support due to the linear convex combination structure in (\ref{mr}) and (\ref{mrib}).} \revisetwo{In numerical experiments, the proposed CML/CMLIB methods demonstrate low estimation bias and consistent performance across multiple settings. While the CML method exhibits greater estimation variability compared to the outcome regression-based approach, we still recommend their use due to their ability to evaluate covariate balance, achieve lower estimation bias, and support tractable statistical inference.}

 By applying the proposed methods, \revisethree{we have estimated the causal effects of alcohol intake frequency on fornix FA, a sensitive marker of brain white matter microstructural integrity. Our findings indicate that daily alcohol intake reduces fornix FA compared with monthly or no alcohol intake. Notable, reduction of fornix FA is clinically meaningful, as lower values have been shown to be associated with age-related white matter degeneration, impaired memory performance, and higher risk of neurodegenerative conditions \citep{kantarci2014fractional, bangen2021decreased} as compared with higher fornix FA values. Recent biological findings suggest that the fornix, a key white matter tract involved in the limbic system and memory consolidation, may be vulnerable to chronic alcohol exposure due to alcohol’s neurotoxic effects on myelination and axonal structure \citep{perez2023alcohol}, which provides biological support for the findings observed in our study. In contrast, our results suggest that \revisetwo{alcohol intake a few times per week may not significantly impact fornix FA}, highlighting a potential threshold effect in which only sustained, heavy alcohol consumption disrupts brain microstructure. These findings are consistent with existing results from human and animal models based on parametric association analyses \citep{zahr2017alcohol, daviet2022associations}. \revisetwo{Additionally, we found that heavy alcohol intake could affect other aging-associated outcomes, including elevated systolic and diastolic measurements \citep{pettersen2014arterial}.} Taken together, our conclusions enrich the existing evidence from a recent clinical study \citep{zhao2023association} and provide additional insights into the relationship between alcohol intake frequency and brain microstructure through a rigorous, quantitative evaluation.} Collectively, these results contribute valuable evidence toward understanding the causal link between alcohol intake and brain health, offering potential guidance for mitigating accelerated neurodegeneration during aging. 

Despite advantages of the proposed estimators, several limitations remain. Although we accounted for a wide range of baseline phenomic variables reflecting health and lifestyle, \revisethree{there may still be unmeasured confounders not captured in our analysis, such as chronic stress or lack of social support that could influence both brain aging and health behaviors.} Moreover, the current CML/CMLIB approaches are limited to \revisetwo{handling multi-categorical exposures. Extending these methods to accommodate continuous exposures, such as alcohol intake measured in grams per week, and time-varying exposures remains an open area for future research.} \revisethree{In addition, several theoretical challenges persist, such as potential heterogeneity between the main and auxiliary datasets and imposing cross-exposure constraints during the calibration step}. These issues are {not fully studied yet in this paper} and warrant further investigation.  


\bibliographystyle{agsm}
\bibliography{Bibliography-MM-MC}

\end{document}